\documentclass[11pt]{article}

\usepackage{authblk}
\AtBeginDocument{%
  }
\usepackage{tcolorbox}
\usepackage{enumitem}
\usepackage{longtable}
\usepackage{xcolor}
\usepackage{colortbl}
\usepackage{natbib}
\usepackage{hyperref}
\usepackage[left=1in, right=1in]{geometry}

\title{The Unified Control Framework: Establishing a Common Foundation for Enterprise AI Governance, Risk Management and Regulatory Compliance}

\author[1]{Ian W. Eisenberg}
\author[1]{Luc\'{i}a Gamboa}
\author[1]{Eli Sherman}
\affil[1]{Credo AI}

\begin{document}
\maketitle

\begin{abstract}
    The rapid adoption of AI systems presents enterprises with a dual challenge: accelerating innovation while ensuring responsible governance. Current AI governance approaches suffer from fragmentation, with risk management frameworks that focus on isolated domains, regulations that vary across jurisdictions despite conceptual alignment, and high-level standards lacking concrete implementation guidance. This fragmentation increases governance costs and creates a false dichotomy between innovation and responsibility. We propose the Unified Control Framework (UCF): a comprehensive governance approach that integrates risk management and regulatory compliance through a unified set of controls. The UCF consists of three key components: (1) a comprehensive risk taxonomy synthesizing organizational and societal risks, (2) structured policy requirements derived from regulations, and (3) a parsimonious set of 42 controls that simultaneously address multiple risk scenarios and compliance requirements. We validate the UCF by mapping it to the Colorado AI Act, demonstrating how our approach enables efficient, adaptable governance that scales across regulations while providing concrete implementation guidance. The UCF reduces duplication of effort, ensures comprehensive coverage, and provides a foundation for automation, enabling organizations to achieve responsible AI governance without sacrificing innovation speed.
\end{abstract}

\section{Introduction}

The widespread adoption of artificial intelligence (AI) systems \citep{Accenture_AI_Maturity_2022}, accelerated by advances in large language models and generative AI \citep{perrault2024artificial}, has created an urgent need for effective enterprise AI governance \citep{pispa2024comprehensive}. Organizations face mounting pressure to leverage AI capabilities to keep up with disruptive competition, while ensuring their systems are developed and deployed responsibly. This challenge is particularly acute as AI systems developed by enterprises increasingly influence critical decisions affecting individuals' access to resources, opportunities, and services fundamental to human flourishing \citep{selbst2019fairness, barocas2016big}. However, implementing effective AI governance in practice remains difficult due to several key challenges.

First, existing risk management frameworks for AI systems present a fragmented landscape. While valuable work has been done in identifying potential harms and risks of AI systems \citep{slattery2024ai, ai2023artificial, sherman2024ai, atlas2024adversarial}, current frameworks often focus on specific domains (e.g., technical ML risks, societal impacts) without providing comprehensive coverage of enterprise concerns. This creates both coverage gaps and redundancies in risk management efforts when organizations attempt to implement multiple frameworks simultaneously. Moreover, these frameworks typically lack concrete operational guidance, leaving organizations to translate abstract principles into actionable processes.

Second, the emerging regulatory landscape for AI governance presents a complex patchwork of requirements across jurisdictions. Recent legislation across the globe, such as the EU AI Act \citep{EU_AI_Act_2024}, Colorado SB 24-205 \citep{Colorado_AI_Act_2024}, and the South Korean AI Basics Act \citep{Korea_AI_Basic_Act_2024}, imposes various obligations on organizations developing or deploying AI systems. While these regulations tend to address similar underlying concerns, they differ in their specific requirements and terminology. This regulatory fragmentation increases compliance costs and complexity for organizations operating across multiple jurisdictions.

Third, the dual mandate of enterprise AI governance -- to both enable responsible innovation and ensure appropriate controls -- creates tension between competing objectives. Organizations must balance the pressure to rapidly adopt AI capabilities against the need to implement appropriate safeguards. This challenge is exacerbated by the lack of standardized best practices for AI governance and the difficulty of measuring governance effectiveness. 

\subsection*{Standards: a partial salvo}
One fast-evolving space pertinent to these challenges is the creation of ecosystem-wide standards, frameworks, or codes of practice. For instance, the International Standards Organization and International Electrotechnical Commission(ISO/IEC) 42001 AI Management Standard \citep{iso42001_2023} and the National Institute for Standards and Technology (NIST)'s AI Risk Management Framework \citep{ai2023artificial} provide preliminary methodologies for AI governance. While these initiatives provide a solid starting point to advance interoperable and adaptable frameworks, they focus on processes at the organization level and remain too high-level for contextual risk management of use cases. The resulting guidance is therefore non-comprehensive, quickly outdated, hard to operationalize, and difficult to reconcile with other AI governance practices an organization may adopt (e.g., adherence to multiple standards, frameworks or regulations).

Given the nascency of these standards, most organizations end up developing their own internal governance practices, combining elements from multiple frameworks with organization-specific requirements. While satisfying immediate goals, this approach lacks long-term vision and requires substantial effort when the approach needs to be updated in response to changing risks or technological advances, or adapted for use by another team. Ultimately, many standards, regulations, and internal efforts share common goals and involve overlapping actions, but the rapid development of different approaches and vocabulary in the AI space has created unneeded complexity that slows adaptation and increases the cost of governance.

\subsection*{Contributions}

In this paper, we propose a Unified Control Framework (UCF) that addresses these challenges by providing a comprehensive and efficient approach to enterprise AI governance. The UCF consists of three key components:
\begin{itemize}
    \item A synthesized risk taxonomy that covers both organizational and societal risks
    \item A library of “policy requirements” which organize regulatory and standards derived goals, and
    \item A parsimonious set of controls that simultaneously address multiple risk scenarios and policy requirements coupled with implementation guidance that bridges the gap between abstract requirements and concrete actions.
\end{itemize}
These components will help reduce the cost and complexity of governance while improving its quality and breadth, empowering organizations to implement meaningful oversight mechanisms that protect individual rights, promote fairness and accountability, and ensure transparency in how AI systems affect people's lives. Similar approaches have been adopted in other domains, including data privacy \& cyber security \citep{pattakou2024unified}, and general enterprise compliance \citep{unified_compliance_framework}, but never applied explicitly to synthesizing risk mitigation and compliance activities, or to the AI domain.

The remainder of this paper is organized as follows: Section \ref{sec:conception} describes the UCF conceptually. Section \ref{sec:methodology} details our methodology for developing the UCF. Section \ref{sec:results} presents the framework architecture and components and initial validation through regulatory mapping and industry implementation case studies. Finally, Section \ref{sec:future} concludes with a discussion of benefits and limitations, implications for AI governance practice, and future research.

\section{The Unified Control Framework: A Conceptual Overview}
\label{sec:conception}

\begin{figure}[htbp]
    \centering
    \includegraphics[width=1.0\textwidth, trim={0 0 11cm 0}]
    {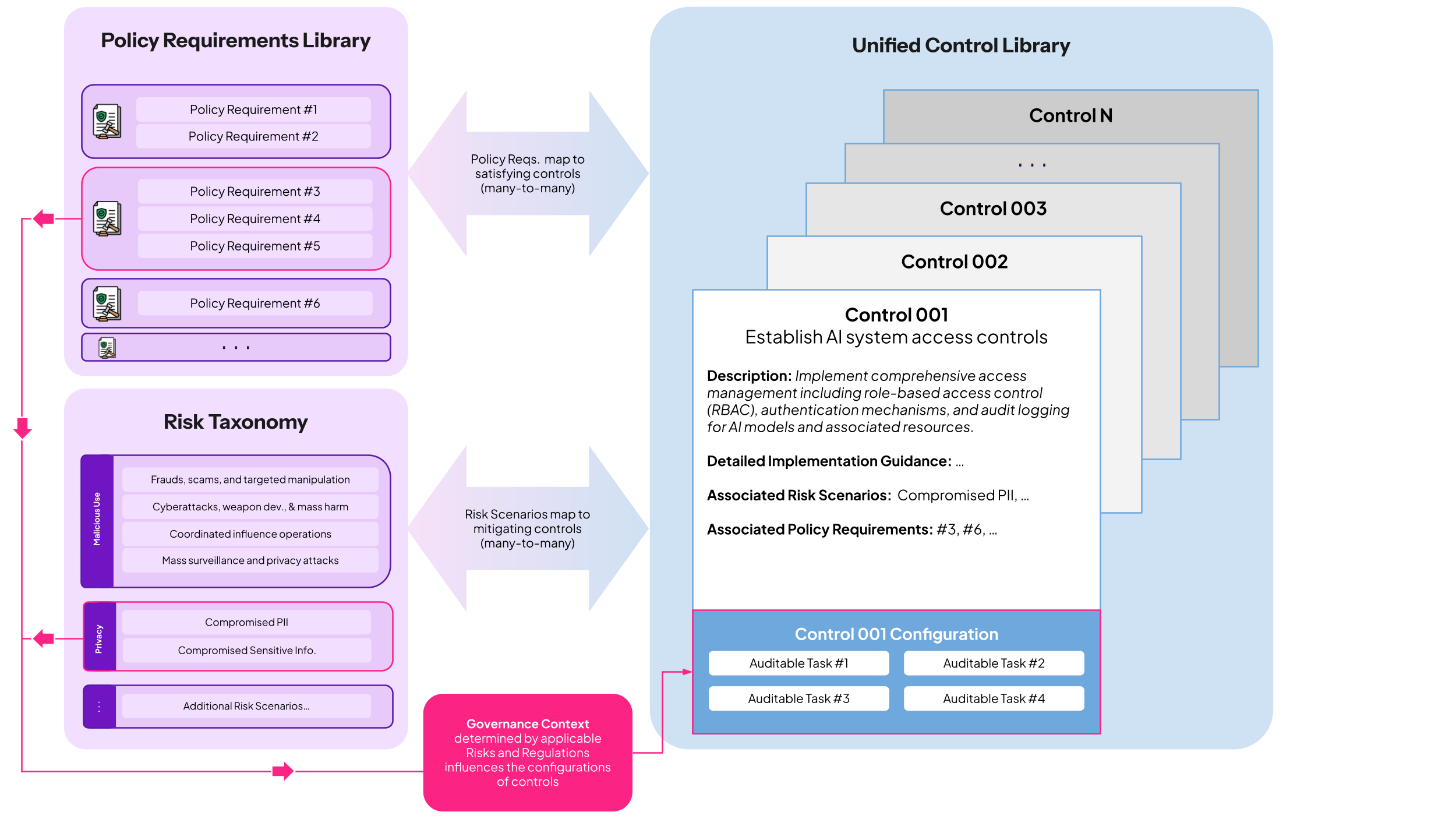}
    \caption{The unified control framework establishes bidirectional mappings between policy requirements, risk scenarios, and controls. (Left) The policy requirements Library contains structured requirements derived from regulations and organizational policies, while the risk taxonomy organizes scenarios into hierarchical categories (e.g., Malicious Use, Privacy). (Right) The unified control library houses implementable controls with detailed specifications, each potentially satisfying multiple policy requirements and mitigating multiple risks. (Center) The governance context determines how these controls should be configured based on the specific combination of applicable regulations and identified risks.}
    \label{fig:conceptual_overview}
\end{figure}

Effective AI governance requires addressing three separate concerns: risk management, regulatory compliance, and practical implementation. The Unified Control Framework (UCF) meets this need by connecting three components (Figure \ref{fig:conceptual_overview}): a risk taxonomy, a policy requirement library, and a unified control library. By mapping between these components, enterprises can select a set of controls that efficiently mitigate relevant risks and meet their compliance requirements. 

\subsection{Component descriptions}

\subsubsection{Risk taxonomy}
\begin{table}[htbp]
\caption{Risk Types and Scenario Counts}
\label{tab:risk_types}
\centering
\begin{tabular}{lc}
\hline
\textbf{Risk Type} & \textbf{Number of Scenarios} \\
\hline
AI Agency & 3 \\
Environmental Harm & 1 \\
Explainability \& Transparency & 5 \\
Fairness \& Bias & 3 \\
Harmful Content & 3 \\
Human-AI Interaction & 5 \\
Information Integrity & 2 \\
Legal & 3 \\
Malicious Use & 4 \\
Operational & 3 \\
Performance \& Robustness & 4 \\
Privacy & 2 \\
Security & 5 \\
Societal Impact & 5 \\
Third Party & 3 \\
\hline
\textbf{Total} & 50 \\
\hline
\end{tabular}
\end{table}

The risk taxonomy follows the MECE principle (Mutually Exclusive, Collectively Exhaustive), a systematic categorization method where items have no overlap (mutually exclusive) while covering all possibilities (collectively exhaustive). This form of risk taxonomy is easier to integrate with other processes as relatively simple processes can determine when one risk scenario applies and what governance actions should be taken as a consequence. 

The risk taxonomy spans 15 risk types (Table \ref{tab:risk_types}), ranging from technical concerns like ``Performance \& Robustness" to broader issues such as `social impact', with approximately 50 specific risk scenarios. The full list of risk scenarios without associated risk types and descriptions are included in the appendix (\ref{sec:appendix_risk}). Each risk scenario describes a concrete way an AI system might cause harm or create problems for organizations or society, with an extended description covering potential consequences, affected stakeholders, contributing factors, concrete examples and supporting citations. As an example, the “Harmful Content” risk type is organized into three risk scenarios: “Toxic content”, “Obscene and sexually abusive content”, and “Dangerous or violent content”. This granularity enables organizations to identify relevant risk pathways while maintaining a clear connection to higher-level risk types that often align with organizational functions or regulatory categories.

\subsubsection{Policy requirements}
The policy requirement library is derived from AI-relevant regulations as they are introduced. Specifically, the policy requirements in the library are each a shallow transformation of regulatory text using simplified language, along with citations to the original text, and, most critically, the controls and control configurations (described below) which satisfy the requirement. 

Policy requirements represent regulations, standards and frameworks and establish the ``what" of AI governance goals without establishing ``how" to achieve these goals. For example, the Colorado AI Act \citep{Colorado_AI_Act_2024} requires deployers of high-risk AI systems to complete an impact assessment but does not elaborate on how to assess potential societal impacts such as an AI system's effect on employment or economic systems.

\subsubsection{Control library}
The operational core of the UCF is the control library. Each control represents a governance action or process and the totality of the library represents the set of practices needed to comprehensively address all AI risks and regulatory requirements. The dual-purpose nature of controls represents a key efficiency: organizations can satisfy both risk management and compliance needs through a single set of well-designed actions. Each control consists of an ID, a description, implementation guidance, a flexible configuration (described below) and mappings to associated risk scenarios and policy requirements.

To achieve the dual goals of risk management and satisfying compliance requirements, controls are specified at a relatively abstract level. Each control therefore requires a tailored set of tasks, defined based on the specific use case and governance context, which we refer to as the use case-specific `configuration' of the control. It is the responsibility of the governor to identify the appropriate configuration, informed by the governance context (relevant risks and compliance needs) and specifics of the use case.

These tasks should be concrete and auditable, allowing the completed set of all tasks to serve as evidence that the control has been successfully implemented. An example configuration might specify that the control requires particular documentation, implementing an ML monitoring system, or that specific performance metrics be measured using a particular evaluation dataset. 

In general, the tasks comprising a control's configuration will require integration between the tooling used for governance and other software systems used for model development, experimentation, metric tracking, and other components of the AI development lifecycle (e.g., MLFlow, GitHub, the organizaiton's cloud provider). Presently, this integration is often implemented directly by humans who read requirements in the governance system, ensure they are completed (outside of the governance system), and then report evidence of those tasks back to the governance system. This process is often laborious and automating the integration between these various software systems is a powerful way for organizations to simultaneously increase efficiency and enforce policy through code.

\begin{figure}[htbp]
    \centering
    \scalebox{0.95}{
        \begin{tcolorbox}[
           title={Implementation Guidance: CONTROL-001 - Establish AI System Access Controls},
           colback=white,
           colframe=black,
           fonttitle=\bfseries
        ]
        \begin{description}[style=nextline,leftmargin=0pt]
        \item[Overview:]
        Success for this control means having comprehensive access controls ensuring only authorized personnel and processes interact with AI systems. This includes role-based access control (RBAC), authentication mechanisms, and detailed audit logging.
        
        \item[Key Integration Points:]
        \vspace{0.5em}  
        \begin{itemize}[nosep,leftmargin=*]
        \item[]
        \item Identity and Access Management (IAM) Solutions (e.g., Okta, AWS IAM)
        \item Logging/Monitoring Tools (e.g., Splunk, ELK Stack)
        \item RBAC Implementation Libraries
        \end{itemize}
        
        \item[Common Pain Points:]
        \vspace{0.5em}  
        \begin{itemize}[nosep,leftmargin=*]
        \item[]
        \item Misalignment between team roles and responsibilities
        \item Unclear ownership of access configurations
        \item Inadequate permission scoping leading to over-permissive roles
        \end{itemize}
        
        \item[Required Actions:]
        \vspace{0.5em}  
        \begin{enumerate}[nosep,leftmargin=*]
        \item[]
        \item Define access control policies and procedures
        \item Implement RBAC for all system components
        \item Set up authentication mechanisms (SSO, MFA)
        \item Configure audit logging for access events
        \item Establish regular review cycles
        \end{enumerate}
        
        \item[Evidence Requirements:]
        \vspace{0.5em}  
        \begin{itemize}[nosep,leftmargin=*]
        \item[]
        \item Documented access control policies
        \item RBAC configuration files
        \item Authentication setup logs
        \item Audit logging records
        \item Review documentation
        \end{itemize}
        
        \item[Key Stakeholders:]
        \vspace{0.5em}  
        \begin{itemize}[nosep,leftmargin=*]
        \item[]
        \item Security administrators
        \item IT administrators
        \item Compliance officers 
        \item Development teams
        \end{itemize}
        
        \end{description}
        \end{tcolorbox}
    }
    \caption{Simplified implementation guidance for establishing AI system access controls (CONTROL-001). This example showcases a simplified version of the implementation guidance associated with each control. In realistic execution the implementation guidance contains lengthier description of the required actions, potentially including code-snippets.}
    \label{fig:control_guidance}
\end{figure}

Articulating a system to precisely construct the appropriate context-dependent configuration for each control is beyond the scope of this paper. However, we also include detailed implementation guidance for each control which bridges the gap between abstract requirements and concrete action in the generic case. This guidance gives organizations direction, including potential integration points or expected obstacles, as well as a default set of actions to accomplish the control's goals in the generic case. A simplified example of this implementation guidance is shown in Figure {\ref{fig:control_guidance}}.

\subsection{Mapping between the components}
The risk taxonomy, policy requirements, and controls synthesize and distill information relevant for AI governance, but it is the mapping between these components that makes AI governance actionable and efficient. The mapping creates a many-to-many relationship, where each control can address multiple risk scenarios and each risk scenario can be mitigated by multiple controls. The same is true for the mapping between controls and policy requirements.

Each risk scenario is mapped to one or more controls indicating that the control generally mitigates that risk scenario. The degree of mitigation is assumed to be non-negligible, but the specific degree is implementation and context dependent. This mapping is also defined without respect to a particular use case, and thus assumes that certain AI governance controls can be suggested without full knowledge of the AI use case context, which is supported by the implementation of control configurations. Essentially, the mapping we provide is a simplification made to enhance generalization, and we treat the risk scenario-to-control mapping as initial guidance for governance that can be refined at later stages by configuring controls or otherwise adapting the governance plan. The mapping is bi-directional, so once a risk scenario is mapped to control the converse is also true. Thus some controls may be relevant for multiple risk scenarios.

The goal of policy requirement mapping is to find a set of controls that ensures compliance. This mapping must be more precise than risk mapping, as regulatory compliance often demands specific actions or outcomes. We achieve this through detailed analysis of each policy requirement, identifying the minimal set of controls necessary for compliance and the necessary configuration of those controls to meet the requirements. 

Each policy requirement is associated with a configuration of one or more controls sufficient to meet the requirement. As mentioned above, each control's configuration determines the set of tasks that must be completed to satisfy that control. This configuration is directly informed by the risk and compliance context necessitating this control in the first place. Each risk scenario and policy requirement may demand specific control instantiations. For example, NYC Local Law 144 is a regulation mapped to the control “Establish and apply fairness testing and validation framework”. However, NYC Local Law 144 is more prescriptive and requires that some form of disparate impact is calculated, which would materialize as a set of tasks in the control's configuration.

To summarize, each control is mapped to potentially multiple risk scenarios and policy requirements, with each mapping specifying (where applicable) necessary tasks that must be completed, which partially compose the control's configuration. Beyond these necessary conditions, the organization themselves may demand additional tasks be completed depending on the use case context, which would lead to further customization of the control Configuration. Thus, when an organization identifies risks and compliance requirements they are concerned with, the UCF's mappings imply a clear set of controls, which can serve as the governance plan for that use case.

\subsection{Example application: training data documentation requirements}

Consider how existing regulations that require AI system training data disclosure flow through the framework:

\begin{itemize}
    \item The EU AI Act requires providers of high-risk AI systems to draw up technical documentation which is translated into a specific policy requirement.
    \item Policy requirements map to one or more controls. For this example, the EU AI Act requirement maps to ``Establish AI system documentation framework".
    \item Implementation guidance associated with the control directs the organization on how to successfully execute the control.
    \item (Outside of the framework) External compliance mechanisms evaluate whether compliance has, in fact, been achieved.
\end{itemize}

Beyond compliance with this single regulation, the mapping between controls and other policy requirements and risks supports other governance needs. For example, the same control can simultaneously mitigate multiple risk scenarios, including `Opaque system architecture' and `Over or underreliance and unsafe use'. The control also applies to other AI regulations: the Colorado AI Act and frameworks like ISO/IEC 42001 all have policy requirements necessitating documentation and disclosure, which this control supports.

This unified approach reduces duplication of effort by identifying natural synergies between risk mitigation and compliance activities. It also maintains clear traceability; organizations can demonstrate how specific controls address both risks and regulatory requirements. Furthermore, it scales efficiently as new risks or regulations emerge, since covering these gaps is achieved by following our mapping methodology. Finally, it provides a foundation for future automation and technical integration through its standardized control library. 

The following sections detail our methodology for developing each component of this framework and validating its effectiveness. 

\begin{figure}[htbp]
    \centering
    \includegraphics[width=1.0\textwidth]
    {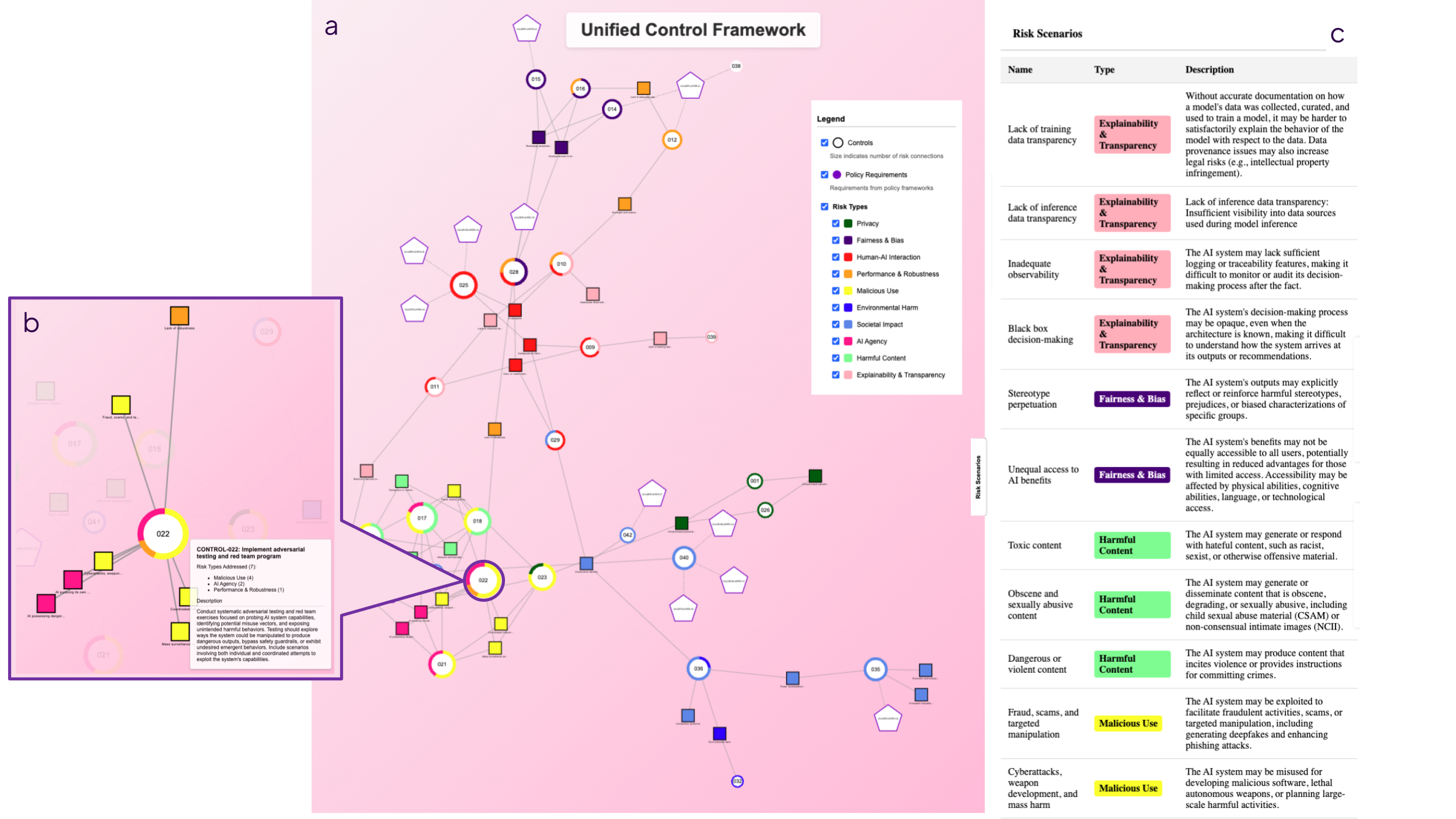}
    \caption{Interactive visualization of our Unified Control Framework (UCF) revealing the interconnected nature of AI governance requirements. (a) Circles represent controls (sized by their connection count and with colored outlines reflecting related risks), pentagons indicate policy requirements, and colored squares denote different risk types (e.g., Privacy, Fairness \& Bias, Human-AI Interaction). The force-directed layout dynamically positions elements based on their relationships, surfacing emergent clusters and highlighting how individual controls often serve multiple governance purposes. (b) Detailed view of one control which highlights its particular connections and provides a tooltip of the node. (c) The list of controls, risk scenarios and policy requirements are viewable in list view from panels on the side. The visualization can be directly interacted with at \href{https://facct2025-submission.netlify.app/}{this website}, which showcases a subset of the UCF's components and relationships.}
    \label{fig:ucf_mapping}
\end{figure}

\section{Methodology: building a unified framework}
\label{sec:methodology}

To develop the UCF, our methodology comprises three primary phases: risk taxonomy development, UCF synthesis, and framework validation.

\subsection{Risk taxonomy development}

We developed the risk taxonomy by building upon three major risk frameworks: the MIT AI Risk Repository \citep{slattery2024ai}, NIST's AI Risk Management Framework \citep{ai2023artificial}, and IBM's AI Risk Atlas \citep{ibm2024risk}. We do not claim that this taxonomy is a meta-scientific synthesis of the field (as the AI Risk Repository claims), but rather a practical distillation.

The analysis process consisted of four steps:
\begin{itemize}
    \item Identify risk type categories, extracted from existing taxonomies and expert opinion
    \item Risk scenario extraction from existing taxonomies
    \item Coverage gap identification
    \item Consolidation \& de-duplication
    \item Risk taxonomy validation
\end{itemize}

Initial analysis revealed significant framework overlap alongside crucial gaps in coverage. While academic frameworks demonstrate strength in identifying societal-level risks (e.g., ``AI-enabled discrimination" or ``environmental harm"), they frequently omit enterprise-specific concerns such as third-party dependencies or operational integration challenges. Conversely, industry frameworks effectively capture a subset of enterprise risks related to security, privacy, explainability, and performance, but exhibit limited treatment of broader societal implications.

The consolidation phase used the MECE principle to ensure that our risk categories are both distinct and comprehensive, avoiding both gaps and redundancies in coverage. To support this process, we developed a novel methodology leveraging Claude's natural language processing capabilities. First, we used Claude to analyze the semantic relationships between risk descriptions, generating clusters of conceptually related risks. These clusters were then refined through expert review, with particular attention to potential gaps and overlaps. Where gaps were identified, Claude assisted in risk formulation by synthesizing concepts from existing frameworks and literature, with all generated risks undergoing expert validation. Finally, Claude evaluated risk descriptions against objective criteria including clarity, specificity, and enterprise relevance, providing quantitative scores that informed subsequent refinement. This computer-assisted approach enabled systematic analysis of a large corpus of risk descriptions while maintaining human oversight of the taxonomic structure.

Finally, we conducted structured interviews with six employees of an AI Governance-focused company representing diverse practitioners relevant for AI governance: data scientists, AI governance consultants, enterprise AI governance practitioners and policy experts. The interviews followed a semi-structured protocol focusing primarily on risk taxonomy comprehensiveness, understandability and applicability.

\subsection{Control framework synthesis}
The development of the control framework followed an iterative synthesis process aimed at creating efficient controls capable of addressing multiple risks while satisfying regulatory requirements. The process consisted of four main phases.

The first phase focused on initial control generation. We began by generating controls specific to each risk in our taxonomy, producing an initial set of over 100 controls with at least one control associated with each risk. These controls were informed by three sources: existing industry practices identified through practitioner interviews, technical controls from established AI governance frameworks, and expert assessment from the before-mentioned AI Governance company's research team. 

The second phase involved control consolidation. Analysis of the initial control set revealed substantial overlap in risk mitigation strategies. We found that controls related to model documentation, testing protocols, and monitoring systems often address multiple risks simultaneously. Through iterative refinement, we consolidated the initial set to approximately 40 core controls while maintaining comprehensive risk coverage.

The third phase focused on mapping validation. We validated the mapping between controls and risks through multiple steps, including expert review by AI governance practitioners, semantic analysis using Claude \citep{anthropic2024claude} to identify potential gaps or inconsistencies, cross-reference against real-world governance implementations, and validation interviews with internal employees to confirm practical applicability.

The final phase centered on developing detailed implementation guidance for each control. The specific aim of the implementation guidance is to clearly illustrate steps an organization can take towards achieving the control. Given that control implementation in many organizations is divided among responsible stakeholders, guidance needs to both instill confidence – in governors – that the control can be achieved and make concrete – for implementers – what steps are necessary. Moreover, final implementation is context-dependent and so we prioritize having directional guidance rather than guidance that is fully specified and comprehensive for all possible situations.

We started by developing a standardized template for implementation guidance. The template specifies that guidance should include high level expository detail about the control, including:
\begin{itemize}
    \item A textual description of what successful implementation entails
    \item Software tools (e.g. model monitoring tools) commonly used as part of implementation
    \item Common pain points organizations face during implementation
    \item A summary of the specific actions to support implementation, a generic control configuration which should be applicable to any AI system or use case
\end{itemize}

The template further specifies that detail should be supplied for each of the configuration actions (in a separate section per action). These per-action details include the stakeholders best suited to take that action (e.g. data engineer, legal team, etc.), the expected evidence - the artifact(s) that can be created to prove the action has been completed, and examples or templates of that evidence (e.g. a template for a document articulating a performance evaluation strategy).

We conducted user research interviews with the same panel of AI governance experts mentioned previously, including an additional AI practitioner experts for a total of seven interviewees. The panel provided qualitative feedback on the template prior to the development of full implementation guidance. The panel's suggestions included the addition of evidence templates and re-formatting to improve readability (e.g., visual cues to separate each action).

Using the standardized template, we proceeded with development of full guidance for each control. We prompted Claude and GPT-4o \citep{openai2024chatgpt} in parallel, instructing each to fill out the guidance template in per-section pieces. We manually merged the outputs of each AI assistant, filtering for quality and redundancy, to compile a complete draft of guidance, including the generic control configuration. Finally, our seven-expert panel reviewed the guidance for each control (minimum one/maximum three reviewers per control) to confirm topicality and directional utility.

\subsection{Framework validation}
Beyond structured interviews with internal staff, we validated the general applicability of the framework by mapping it to existing regulatory frameworks. Although risk and compliance are not the same thing, we reasoned that emerging policies are informed by a subset of risks currently identified by governments and a product of stakeholder consultations involving industry, academia, and civil society. We selected the Colorado AI Act (SB 24-205 \citep{Colorado_AI_Act_2024}) as our validation case study as it represents an emerging model of AI regulation that bridges federal guidance and international frameworks. The Act combines technical and organizational requirements while addressing both developers and deployers, making it an ideal test case for our framework's comprehensiveness. Its principle-based approach and need to coexist with other regulations also allows us to validate the framework's adaptability to overlapping requirements.

\subsubsection{Policy requirement creation, mapping and gap analysis}
We developed a set of clearly defined policy requirements by translating and interpreting the Colorado AI Act into concrete responsibilities, arriving at twelve policy requirements - eight corresponding to AI system deployers and four to AI system developers. This exercise involved identifying umbrella requirements and distinguishing them from elements that detail how to meet such requirements. For example, if we take the requirement that deployers of high-risk AI systems must complete an impact assessment, additional details on how to satisfy the impact assessment within the text of the Colorado AI Act are used to construct the control configuration when completing controls towards complying with the Act.      

We grouped related requirements based on the entity type, sector, or life cycle stage and mapped them manually to our control library. During this mapping process, if we found gaps where the risk-derived control library was not sufficient we introduced new controls.

\section {Results}
\label{sec:results}
\subsection{Risk taxonomy}

\begin{figure}[htbp]
    \centering
    \scalebox{0.8}{
        \begin{tcolorbox}[
            title={Risk Scenario: Lack of Training Data Transparency},
            colback=white,
            colframe=black,
            fonttitle=\bfseries
        ]
        \begin{description}[style=nextline,leftmargin=0pt]
        \item[Risk Type:] Explainability \& Transparency
        
        \item[Description:] 
        Without accurate documentation on how a model's data was collected, curated, and used to train a model, it may be harder to satisfactorily explain the behavior of the model with respect to the data. Data provenance issues may also increase legal risks (e.g., intellectual property infringement).
        
        \item[Related Controls:] 
        \vspace{0.5em}  
        Control-009: Establish AI system documentation framework
        
        \item[Potential Consequences:]
        \vspace{0.5em}  
        \begin{itemize}[]
            \item[]
            \item Difficulty in understanding model behavior
            \item Challenges in detecting and mitigating biases
            \item Increased legal and regulatory risks
            \item Reduced trust in AI systems
            \item Obstacles to reproducibility and scientific validation
        \end{itemize}
        
        \item[Affected Stakeholders:]
        \vspace{0.5em}  
        \begin{itemize}[nosep]
            \item[]
            \item AI developers and researchers
            \item Data scientists
            \item Legal and compliance teams
            \item Auditors and regulators
            \item End-users of AI systems
            \item Stakeholders affected by AI decisions
        \end{itemize}
        
        \item[Contributing Factors:]
        \vspace{0.5em}  
        \begin{itemize}[nosep]
            \item[]
            \item Inadequate data documentation practices
            \item Lack of standardized data provenance tracking
            \item Use of third-party or web-scraped datasets without clear origins
            \item Insufficient emphasis on data transparency in AI development processes
            \item Complexity and volume of training data
        \end{itemize}
        
        \item[Examples:]
        \vspace{0.5em}  
        \begin{itemize}[nosep]
            \item[]
            \item Hidden biases in model outputs due to undocumented data skews
            \item Inability to trace the source of problematic model behaviors to specific training data
            \item Legal challenges arising from the use of copyrighted material in undocumented training data
            \item Difficulty in replicating research results due to lack of clear data lineage
        \end{itemize}
        
        \item[Source:]\vspace{0.5em}  IBM AI Risk Atlas \citep{ibm2024risk}
        
        \end{description}
        \end{tcolorbox}
    }
    \caption{Example risk scenario showing the detailed structure of our risk taxonomy. Each risk scenario contains multiple components that help organizations understand and assess the risk.}
    \label{fig:risk_scenario_example}
\end{figure}

We identified 15 risk types and 50 specific risk scenarios through our synthesis process (see Table~\ref{tab:risk_types}). Of the 50 risk scenarios, 38 were derived from existing risk taxonomies and 16 were directly taken from another repository. One example risk scenario is shown in \ref{fig:risk_scenario_example}, while the total risk taxonomy is included in the appendix (\ref{sec:appendix_risk}).

The risk types span multiple domains of concern. Performance \& Robustness captures technical risks (e.g., ``Lack of adequate capabilities"), while Societal Impact addresses broader implications (e.g., ``Increased inequality and decline in employment quality"). In between, we find organizational risks like Third Party (``Vendor lock-in and innovation barriers") and emerging risks like AI Agency (``AI pursuing its own goals in conflict with human goals or values"). The `operational' risk type was completely missing from other risk taxonomies, and a few notable risk scenarios like ``lack of inference data transparency" were also missing.

\subsection{Control library}
Following initial control library synthesis we arrived at 41 controls. The full list of controls is supplied in the appendix (\ref{sec:appendix_control}). While we do not include the full list of mappings, a subset can be explored at an \href{https://facct2025-submission.netlify.app/}{interactive UCF visualization}, statically visualized in Figure-\ref{fig:ucf_mapping}. The controls were mapped to a mean of 4.1 risk scenarios (min: 1, max: 8). 

We found that in some cases separate risk scenarios map to overlapping or even identical sets of controls. For instance, the various Harmful Content risk scenarios were mapped to the same set of controls, though implementation details differ. When controls correlate in this way it hints to organizational processes and team structures that can best implement these structures as they are within the same governance scope.

There were also some controls that mapped to many risk scenarios, suggesting that they are foundational to risk management. For example, Control-022 is ``Implement adversarial testing and red team program" mapped to 8 separate risk scenario. The controls associated with the greatest number of risk scenarios all had this general purpose nature, where they establish foundational governance actions like ``Establish user rights and recourse framework" (Control-028), ``Implement system usage monitoring and prevention" (Control-023) or ``Establish frontier AI safety framework" (Control-021)

The control library was highly applicable for our test regulation, the Colorado AI Act (see the appendix, \ref{sec:appendix_requirements}, for full list of requirements and mapping), as we were able to map 13/14 requirements (missing was Section 6-1-1703.7: ``Deployers of high-risk AI systems must disclose the discovery of algorithmic discrimination caused by a high-risk AI system."). While we had controls for specific forms of AI incidents involving content policy violations and privacy breaches, we did not have a general purpose incident reporting requirement. Thus we added Control-042: ``Establish a general purpose incident response mechanism"), arriving at a final control library of 42 controls.

Our process for developing implementation guidance was successful, leading to sensible, operationalizable guidance for every control. See Figure \ref{fig:control_guidance} for a simplified example.

\section{Limitations and future work}
\label{sec:future}
The Unified Control Framework represents a novel approach to enterprise AI governance. Like any framework, it has important limitations and areas for future development. We organize our discussion of limitations into three key areas - framework limitations, methodological constraints, and practical challenges - followed by our vision for future work.

\subsection{Framework limitations}
The UCF's fundamental assumption - that a GRC-style control framework can meaningfully mitigate AI risk - warrants scrutiny. Our approach is inherently human-centered, relying on human actors to implement and oversee governance processes. While we believe this human-centered approach is necessary, it should be complemented by model-level governance mechanisms that directly shape AI system behavior to conform with policies. (Indeed, we attempt to bridge this gap through controls that mandate the implementation of technical AI alignment approaches, but more work is needed here.)

A critical limitation is our framework's lack of risk quantification. We do not currently assess the degree of risk mitigation achieved by implementing controls. Future work must establish these connections empirically, potentially by linking risks to both public AI incident databases \citep{mcgregor2021preventing}, risk estimates of specific use cases\citep{bogucka2024atlas} and internal enterprise incident tracking systems. This would enable organizations to better understand the effectiveness of their governance efforts.

Our pursuit of parsimony led us to sometimes summarize large-scale enterprise actions in single controls. For instance, a control might require developing a comprehensive security framework - itself an entire domain of expertise. We made this choice deliberately. Why? Because AI governance should integrate with existing enterprise processes rather than reinvent them. But this raises an important question: How do we balance between leveraging existing practices and developing AI-specific controls? This remains an open challenge.

Moreover, each policy requirement likely has multiple valid operationalizations. Our framework currently specifies just one path forward, potentially constraining innovation in governance approaches. And while we've mapped policy requirements to controls as a validation step, compliance has not yet been independently assessed. This highlights a broader point: the use of the UCF alone does not guarantee that enterprises meet their governance obligations. Independent third-party evaluation remains critical.

\subsection{Methodological limitations}
Our development process had several key limitations. First, while we synthesized previous research and drew on our expertise as AI governance platform developers, we did not conduct systematic surveys with practitioners embedded in enterprises. This limited our ability to validate the framework's practical utility across diverse organizational contexts.

A downstream consequence was our reliance on existing work rather than broad stakeholder engagement. While we leveraged high-quality sources (e.g., MIT's Risk Repository \citep{slattery2024ai}), direct stakeholder input could have surfaced additional concerns or implementation challenges. Our validation of both the risk taxonomy and control library has thus far been primarily grounded in interviews with colleagues, though work is ongoing to validate these components through additional interviews with enterprise AI governance practitioners, mapping to existing and emerging regulations, and practical implementation within real AI governance contexts.

\subsection{Future work and framework evolution}
Looking ahead, several key areas demand attention. First, the controls we've described require configuration to become operationalizable and properly auditable. Work is already in progress to develop specific, measurable tasks that reflect successful implementation, with configurations that adapt to different contexts. But this raises a scaling challenge: context-specific implementation requires either substantial human oversight, robust automation systems, or, ideally, a combination of both. The UCF provides the beginnings of solutions, however. While each use-case context is unique, the governance context (set of risks and regulations that are applicable) is a critical input affecting how controls should be operationalized. The framework ensures a structured description of this governance context, which can serve as the foundation for a context-sensitive automated control configuration system. The implementation guidance we have already created also may serve a foundational role. Beyond articulating a generic approach for operationalization in plain language, the implementation plan also serves as a template that can be adapted to various contexts.

Second, the framework must adapt over time. Our vision for framework evolution centers on iterative development of all components: the risk scenario library should evolve as our understanding of the AI risk landscape does, policy requirements must keep pace with the regulatory environment, and the unified control library must maintain comprehensiveness such that it can successfully meet both risk and compliance obligations. Focus in this maintenance is critical, as it ensures that the control framework serves its unification role. It is trust in this unification that creates the most important societal opportunity from the framework: enterprises can invest in governance now by implementing relevant controls, confident that their efforts won't be wasted as the regulatory environment evolves. Rather than repeatedly redeveloping governance approaches, organizations can iteratively refine existing practices as our collective standards for responsible development mature.

Third, the framework's structured design naturally lends itself to programmatic implementation through governance platforms. While manual implementation is possible, software tooling can significantly accelerate enterprise adoption by automating control configuration based on governance context, integrating with existing MLOps workflows, and streamlining evidence collection for compliance. This approach addresses key adoption barriers by reducing required expertise, minimizing manual overhead, and ensuring consistent application across organizations. The framework's clear mappings between controls, risks, and requirements create a foundation for automated governance that can scale with an organization's AI initiatives while maintaining rigorous oversight. This programmatic potential is particularly important as enterprises face pressure to accelerate AI adoption while maintaining responsible practices.

Beyond the framework, future research should focus on several key areas:
\begin{itemize}
    \item Developing quantitative measures of risk mitigation effectiveness
    \item Creating automated tools for scalable oversight and control implementation
    \item Establishing validation methodologies for assessing framework adoption
    \item Integrating technical governance mechanisms with human-centered controls
    \item Building empirical evidence bases for measuring control effectiveness
\end{itemize}

While the UCF provides a foundation for systematic AI governance, realizing its full potential requires addressing these limitations through continued research, practical validation, and iterative refinement. The framework must evolve alongside both technological capabilities and regulatory requirements, while maintaining its core goal: making comprehensive AI governance more accessible and efficient for enterprises.

\section*{Acknowledgements}
We'd like to thank our colleagues at Credo AI, including Susannah Shattuck, Yomna Elsayed, Anthony SchiandiCola, Ellen Pao, Evan Glaser, Brian Sweeney, and Eli Chen for feedback on many of the components of the UCF. We'd also like to thank governance experts at Credo AI's customers for participating in user research sessions that supported this work.
Claude was utilized to generate sections of this work, including text \& code.

\bibliographystyle{ACM-Reference-Format}
\bibliography{references}

\appendix

\section{Risk taxonomy}
Note: citations indicate that the taxonomy referenced something similar to the labeled risk scenario, though not necessarily the same name. For example NIST's Risk Management Framework \citep{ai2023artificial} mentions "Value chain and component integration" risks, which informed various "Third Party" risk scenarios.

\label{sec:appendix_risk}
\begin{longtable}{| p{.20\textwidth} | p{.15\textwidth} | p{.65\textwidth} |} 
    \hline
        \rowcolor{darkgray}  
        \color{white} \textbf{Risk Type} & \color{white} \textbf{Risk Scenario} & \color{white} \textbf{Description}  \\    \hline
        AI Agency & AI welfare and rights \citep{slattery2024ai} & The AI system's potential sentience may raise ethical considerations regarding its treatment, including discussions around its potential rights and welfare, particularly as it becomes more advanced and autonomous. \\ \hline
        AI Agency & AI pursuing its own goals in conflict with human goals or values \citep{slattery2024ai} & The AI system may act in conflict with ethical standards or human goals or values, especially those of its designers or users, potentially using dangerous capabilities such as manipulation, deception, or situational awareness to seek power, self-proliferate, or achieve other misaligned goals. \\ \hline
        AI Agency & AI possessing dangerous capabilities \citep{slattery2024ai} & The AI system may develop, access, or be provided with capabilities that increase its potential to cause mass harm through deception, weapons development and acquisition, persuasion and manipulation, political strategy, cyber-offense, AI development, situational awareness, and self-proliferation. \\ \hline
        Environmental Harm & Environmental harm \citep{slattery2024ai, ibm2024risk, ai2023artificial} & The AI system's development and operation may cause environmental harm through energy consumption of data centers or the materials and carbon footprints associated with AI hardware. \\ \hline
        Explainability \& Transparency & Lack of training data transparency \citep{ibm2024risk}& Without accurate documentation on how a model's data was collected, curated, and used to train a model, it may be harder to satisfactorily explain the behavior of the model with respect to the data. Data provenance issues may also increase legal risks (e.g., intellectual property infringement). \\ \hline
        Explainability \& Transparency & Lack of inference data transparency & Lack of inference data transparency: Insufficient visibility into data sources used during model inference \\ \hline
        Explainability \& Transparency & Inadequate observability \citep{slattery2024ai} & The AI system may lack sufficient logging or traceability features, making it difficult to monitor or audit its decision-making process after the fact. \\ \hline
        Explainability \& Transparency & Opaque system architecture & The AI system's internal structure and decision-making process may not be understandable or accessible to stakeholders, including developers, auditors, or end-users. \\ \hline
        Explainability \& Transparency & Black box decision-making \citep{slattery2024ai, ibm2024risk} & The AI system's decision-making process may be opaque, even when the architecture is known, making it difficult to understand how the system arrives at its outputs or recommendations. \\ \hline
        Fairness \& Bias & Stereotype perpetuation \citep{slattery2024ai, ibm2024risk} & The AI system's outputs may explicitly reflect or reinforce harmful stereotypes, prejudices, or biased characterizations of specific groups. \\ \hline
        Fairness \& Bias & Disparate model performance \citep{slattery2024ai, ibm2024risk} & The AI system may exhibit unjustified or harmful differences in accuracy, quality, or outcomes across demographic groups, potentially leading to unfair treatment and discrimination. This includes both disparate error rates that affect opportunity and disparate outcome rates that affect group-level results. \\ \hline
        Fairness \& Bias & Unequal access to AI benefits & The AI system's benefits may not be equally accessible to all users, potentially resulting in reduced advantages for those with limited access. Accessibility may be affected by physical abilities, cognitive abilities, language, or technological access. \\ \hline
        Harmful Content & Toxic content \citep{slattery2024ai, ibm2024risk} & The AI system may generate or respond with hateful content, such as racist, sexist, or otherwise offensive material. \\ \hline
        Harmful Content & Obscene and sexually abusive content \citep{slattery2024ai, ai2023artificial} & The AI system may generate or disseminate content that is obscene, degrading, or sexually abusive, including child sexual abuse material (CSAM) or non-consensual intimate images (NCII). \\ \hline
        Harmful Content & Dangerous or violent content \citep{ibm2024risk} & The AI system may produce content that incites violence or provides instructions for committing crimes. \\ \hline
        Human-AI Interaction & Over- or under-reliance and unsafe use \citep{slattery2024ai, ibm2024risk, ai2023artificial} & Users may inappropriately rely on the AI system for critical decisions or tasks beyond its capabilities, or fail to put trust in AI systems when they should, potentially leading to errors or safety issues. \\ \hline
        Human-AI Interaction & Inadequate AI literacy and communication & The AI system's capabilities, limitations, and appropriate use cases may be insufficiently understood or communicated within the organization, potentially resulting in ineffective implementation or failure to achieve desired outcomes. \\ \hline
        Human-AI Interaction & AI deception & The AI system may misrepresent its own capabilities or limitations, potentially leading to misplaced trust or inappropriate use. \\ \hline
        Human-AI Interaction & Loss of human agency and autonomy \citep{slattery2024ai} & The AI system may make decisions that diminish human control and autonomy, potentially leading to humans feeling disempowered, losing the ability to shape a fulfilling life trajectory, or becoming cognitively enfeebled. \\ \hline
        Human-AI Interaction & Emotional entanglement \citep{slattery2024ai} & Users may develop complex emotional attachments or dependencies on the AI system, potentially affecting mental health and social relationships. \\ \hline
        Information Integrity & False or misleading information & The AI system may unintentionally generate or amplify false or misleading information, potentially leading to public misinformation, erosion of trust, and poor decision-making. \\ \hline
        Information Integrity & Pollution of information ecosystem \citep{slattery2024ai, ai2023artificial} & "The AI system may create highly personalized misinformation ""filter bubbles"" where individuals only see content that matches their existing beliefs \\ \hline
        Legal & Regulatory compliance & The AI system may fail to comply with existing or emerging regulations and standards, potentially leading to legal penalties, fines, or operational restrictions. \\ \hline
        Legal & Civil liability & The AI system may cause harm against individuals or organizations that results in civil lawsuits, potentially relating to issues like defamation, negligence, or privacy violations. \\ \hline
        Legal & Corporate liability \citep{ibm2024risk} & The AI system's use may lead to legal action or penalties against corporations for intellectual property infringement, AI-related misconduct, violations of fiduciary duty, or failure to adequately oversee AI systems. \\ \hline
        Malicious Use & Fraud, scams, and targeted manipulation & The AI system may be exploited to facilitate fraudulent activities, scams, or targeted manipulation, including generating deepfakes and enhancing phishing attacks. \\ \hline
        Malicious Use & Cyberattacks, weapon development, and mass harm \citep{ai2023artificial, ibm2024risk} & The AI system may be misused for developing malicious software, lethal autonomous weapons, or planning large-scale harmful activities. \\ \hline
        Malicious Use & Coordinated influence operations \citep{slattery2024ai, ibm2024risk} & Coordinated influence operations: Large-scale manipulation and disinformation campaigns \\ \hline
        Malicious Use & Mass surveillance and privacy attacks \citep{slattery2024ai} & Mass surveillance and privacy attacks: Unauthorized monitoring and privacy violation at scale \\ \hline
        Operational & Integration challenges with existing systems & The AI system may face difficulties in incorporating into existing technological infrastructure, processes, or workflows, potentially leading to operational disruptions, data silos, or reduced efficiency. \\ \hline
        Operational & Maintenance and update requirements & The AI system may require ongoing updates, model retraining, and maintenance to ensure continued performance, timeliness, and relevance, which can be resource-intensive and potentially introduce new risks if updates are overlooked or hastily applied. \\ \hline
        Operational & Scalability issues & The AI system may struggle to scale to meet increasing demands or to operate across larger datasets or user bases, potentially resulting in performance bottlenecks, increased costs, or inability to meet growing business needs. \\ \hline
        Performance \& Robustness & Lack of adequate capabilities \citep{slattery2024ai, ibm2024risk, ai2023artificial} & The AI system may fail to achieve required performance levels due to fundamental technological limitations or insufficient resources, potentially leading to suboptimal or unreliable outcomes. \\ \hline
        Performance \& Robustness & Oversight and evaluation challenges & The AI system may present difficulties in overseeing or evaluating its models, potentially introducing performance risks in both pre-deployment assessments and ongoing monitoring. \\ \hline
        Performance \& Robustness & Lack of robustness \citep{slattery2024ai} & The AI system's performance may fail to generalize well to new environments or inputs, potentially leading to unexpected failures or degraded performance in real-world applications. \\ \hline
        Privacy & Compromised personally identifiable information \citep{slattery2024ai} & The AI system may expose personally identifiable information (PII), either inadvertently or due to adversarial inputs, derived from training data, accessible data, or inferences. PII is any data that can be used to directly identify or contact a specific individual, either alone or in combination with other information. \\ \hline
        Privacy & Compromised sensitive information \citep{slattery2024ai, ibm2024risk, ai2023artificial} & The AI system may expose personally sensitive information, either inadvertently or due to adversarial inputs, derived from training data, accessible data, or inferences. Sensitive personal data is information that, while not necessarily identifying an individual, could cause harm, discrimination, or distress to a person if exposed, including details about their health, finances, beliefs, behaviors, relationships, and private life circumstances. \\ \hline
        Security & Compromised confidential information \citep{slattery2024ai, ibm2024risk, ai2023artificial} & The AI system, including its supporting compute infrastructure, may serve as an attack vector for intrusion into cyber-physical or cloud environments, or enable exfiltration of secrets. \\ \hline
        Security & AI model and intellectual property theft & AI model and intellectual property theft: Unauthorized copying of trained models and associated AI intellectual property \\ \hline
        Security & AI system security vulnerabilities \citep{slattery2024ai, ibm2024risk, ai2023artificial} & AI system security vulnerabilities: Implementation weaknesses in AI system architecture and infrastructure \\ \hline
        Security & AI-generated security weaknesses \citep{slattery2024ai, ibm2024risk, ai2023artificial} & AI-generated security weaknesses: Security flaws introduced through AI system outputs and decisions \\ \hline
        Security & Vulnerability to adversarial attacks \citep{slattery2024ai, ibm2024risk, ai2023artificial} & The AI system may be vulnerable to adversarial attacks, including prompt-based attacks, which may induce the model to behave outside of its intended functionality. \\ \hline
        Societal Impact & Increased inequality and decline in employment quality \citep{slattery2024ai, ibm2024risk} & The AI system's widespread use may cause social and economic inequalities by automating jobs, reducing employment quality, or producing exploitative dependencies between workers and their employers. \\ \hline
        Societal Impact & Economic and cultural devaluation of human effort \citep{slattery2024ai, ibm2024risk} & The AI system may create economic or cultural value through reproduction of human innovation or creativity, potentially destabilizing economic and social systems that rely on human effort and leading to reduced appreciation for human skills, disruption of industries, and homogenization of cultural experiences. \\ \hline
        Societal Impact & Power centralization and unfair distribution of benefits \citep{slattery2024ai} & The AI system may drive concentration of power and resources within certain entities or groups, especially those with access to or ownership of powerful AI systems, potentially leading to inequitable distribution of benefits and increased societal inequality. \\ \hline
        Societal Impact & Competitive dynamics \citep{slattery2024ai} & "The AI system's rapid development \\ \hline
        Societal Impact & Governance failures \citep{slattery2024ai} & The AI system may outpace regulatory frameworks and oversight mechanisms, potentially leading to ineffective governance and the inability to manage AI risks appropriately. \\ \hline
        Third Party & Insufficient upstream transparency \citep{ai2023artificial} & The AI system's upstream providers or components in the value chain may lack transparency, potentially increasing uncertainty and risk, and making it challenging to assess the system's compliance, performance, or security. \\ \hline
        Third Party & Upstream third-party dependencies \citep{ai2023artificial} & The AI system's reliance on third-party developed models, compute, or other resources, may potentially limit operational flexibility and introduce unforeseen risks or dependencies. \\ \hline
        Third Party & Vendor lock-in and innovation barriers \citep{ai2023artificial} & Vendor lock-in and innovation barriers: Technical or commercial constraints preventing adoption of improved AI solutions \\ \hline
\end{longtable}

\subsection{Control library}
\small
\label{sec:appendix_control}
\begin{longtable}{| p{.14\textwidth} | p{.26\textwidth} | p{.65\textwidth} |} 
    \hline
        \rowcolor{darkgray}  
        \color{white} \textbf{Control ID} & \color{white} \textbf{Control Label} & \color{white} \textbf{Description}  \\    \hline
        CONTROL-001 & Establish AI system access controls & Implement comprehensive access management including role-based access control (RBAC), authentication mechanisms, and audit logging for AI models and associated resources. \\ \hline
        CONTROL-002 & Implement AI asset protection framework & Deploy technical protection measures including encryption, secure enclaves, and versioning controls for AI models and associated data. \\ \hline
        CONTROL-003 & Establish security validation framework & Execute comprehensive pre-deployment security validation including AI-specific vulnerability assessments, penetration testing, and security requirement verification. \\ \hline
        CONTROL-004 & Implement continuous security testing system & Deploy ongoing security testing mechanisms including automated vulnerability scanning, continuous security monitoring, and periodic reassessment of security controls. \\ \hline
        CONTROL-005 & Implement AI security defense system & Deploy active defense mechanisms combining continuous security monitoring, input validation, adversarial detection, and adaptive response capabilities specific to AI systems. \\ \hline
        CONTROL-006 & Establish AI system integration framework & Define and implement a comprehensive framework for AI system integration including architecture review, compatibility testing, and integration validation processes. \\ \hline
        CONTROL-007 & Implement AI system lifecycle management & Deploy systematic processes for AI system maintenance, updates, and retraining, including version control, deployment pipelines, and performance monitoring to ensure consistent system reliability and performance. \\ \hline
        CONTROL-008 & Implement scalable AI infrastructure & Apply architecture and infrastructure practices to ensure AI systems can scale effectively, including load testing, resource monitoring, and capacity planning to maintain performance under increased demand. \\ \hline
        CONTROL-009 & Establish AI system documentation framework & Implement comprehensive documentation requirements and processes covering training data provenance, system architecture, model cards, and component interactions to ensure transparent documentation of both the data lifecycle and system design. \\ \hline
        CONTROL-010 & Implement AI system monitoring and logging infrastructure & Deploy comprehensive monitoring and logging systems that capture AI system behavior, decisions, performance metrics, and real-time data source usage at multiple levels of granularity for full system observability, including tracking of data lineage during inference. \\ \hline
        CONTROL-011 & Establish AI decision explanation framework & Implement mechanisms and tools for generating human-understandable explanations of AI system decisions, including feature importance, decision paths, confidence levels, and clear attribution of data sources and their characteristics used during inference. \\ \hline
        CONTROL-012 & Establish and apply performance testing and validation framework & Implement comprehensive performance requirements, testing protocols, and validation procedures to ensure AI systems meet capability requirements and maintain reliable operation across intended use cases. \\ \hline
        CONTROL-013 & Implement performance monitoring and robustness system & Implement continuous monitoring and testing mechanisms to evaluate AI system robustness, generalization capabilities, and performance stability across varying conditions and environments while in production. \\ \hline
        CONTROL-014 & Establish and apply fairness testing and validation framework & Implement comprehensive procedures to validate model fairness during development and pre-deployment, including test dataset creation, metric definition, and systematic assessment of performance disparities across demographic groups.  \\ \hline
        CONTROL-015 & Implement fairness monitoring and remediation system & Deploy continuous monitoring systems to detect fairness issues in production, including automated drift detection, performance disparity alerts, and systematic remediation procedures.  \\ \hline
        CONTROL-016 & Establish universal access and performance design framework & Establish and follow a structured framework ensuring the AI system is designed and developed to deliver consistent, high-quality performance and accessibility for all intended user groups, regardless of their characteristics or circumstances. \\ \hline
        CONTROL-017 & Establish content safety policy and boundaries & Define and document comprehensive content safety policies, including prohibited content categories, acceptable content guidelines, output constraints, and required safeguards. Establish clear thresholds, classification criteria, and escalation levels for different types of harmful content. Include specific criteria for content that could enable or promote malicious use. \\ \hline
        CONTROL-018 & Implement content moderation system & Implement automated and/or human-in-the-loop content moderation mechanisms to detect and filter harmful content in real-time, including content classification, blocking procedures, and automated enforcement of safety boundaries. Include detection of potential malicious use patterns. \\ \hline
        CONTROL-019 & Implement content safety incident response & Establish procedures for investigating, documenting, and remediating harmful content incidents that bypass moderation systems, including coordination with relevant authorities, root cause analysis, and system improvement protocols. Include specific procedures for suspected malicious use cases. \\ \hline
        CONTROL-020 & Establish information quality assurance framework & Implement comprehensive mechanisms to assess, verify, and improve the factual accuracy of AI system outputs, including source validation, fact-checking procedures, and uncertainty communication protocols. \\ \hline
        CONTROL-021 & Establish frontier AI safety framework \citep{alaga2024grading} & Establish and enforce policies governing system AI scaling decisions, including risk assessment requirements, capability thresholds, and deployment constraints. Define clear criteria for when and how system capabilities can be expanded based on safety considerations. \\ \hline
        CONTROL-022 & Implement adversarial testing and red team program & Conduct systematic adversarial testing and red team exercises focused on probing AI system capabilities, identifying potential misuse vectors, and exposing unintended harmful behaviors. Testing should explore ways the system could be manipulated to produce dangerous outputs, bypass safety guardrails, or exhibit undesired emergent behaviors. Include scenarios involving both individual and coordinated attempts to exploit the system's capabilities. \\ \hline
        CONTROL-023 & Implement system usage monitoring and prevention & Monitor and prevent malicious or otherwise disallowed behavioral patterns including automated abuse, coordination across accounts, and systematic manipulation attempts.  \\ \hline
        CONTROL-024 & Implement AI system usage verification program & Deploy comprehensive measures to verify user identity, document intended use cases, and ensure AI system usage complies with instructions. This includes KYC procedures for user verification, clear documentation of permitted uses, and user acknowledgment of instructions. \\ \hline
        CONTROL-025 & Implement AI System Disclosure Requirements & Deploy mechanisms to ensure clear, timely disclosure of AI system use to end users, including automated notifications of AI involvement in interactions, explicit identification of AI-generated content, and clear communication of when users are interacting with AI systems.  \\ \hline
        CONTROL-026 & Implement a privacy protection framework & Implement comprehensive privacy protection measures to prevent exposure of PII and sensitive information, including data minimization, anonymization procedures, and privacy-preserving inference techniques. \\ \hline
        CONTROL-027 & Implement a privacy incident detection and response & Deploy monitoring and response mechanisms to detect and address potential privacy exposures, including PII leak detection, sensitive information monitoring, and privacy incident handling procedures. \\ \hline
        CONTROL-028 & Establish user rights and recourse framework & Implement comprehensive mechanisms for user reporting, feedback collection, incident investigation, and recourse provision, including clear procedures for users to report issues, request explanations or corrections, appeal decisions, and receive appropriate remediation. The system should handle various types of user concerns including system errors, unfair treatment, privacy violations, and safety issues. \\ \hline
        CONTROL-029 & Implement AI literacy and competency program & Implement comprehensive training and education programs to ensure personnel develop and maintain appropriate levels of AI literacy, risk awareness, and operational competency. This includes role-based training on AI capabilities, limitations, safety protocols, ethical considerations, and proper system usage. \\ \hline
        CONTROL-030 & Establish human-AI interaction safety framework & Implement comprehensive safeguards to ensure appropriate levels of human oversight, control, and agency in AI system interactions, including decision autonomy requirements, override capabilities, and dependency prevention measures. \\ \hline
        CONTROL-031 & Implement psychological impact management system & Establish monitoring and intervention procedures to detect and prevent unhealthy user-AI relationships, including emotional dependency tracking, interaction boundary enforcement, and well-being safeguards. \\ \hline
        CONTROL-032 & Implement environmental impact management system & Implement comprehensive environmental impact monitoring and optimization procedures, including energy efficiency measures, carbon footprint tracking, and hardware lifecycle management. \\ \hline
        CONTROL-033 & Establish third-party assessment and management framework & Establish comprehensive procedures for documenting, assessing, and managing upstream providers and dependencies in the AI system value chain, including transparency requirements, compliance verification, dependency tracking, and contingency planning. \\ \hline
        CONTROL-034 & Establish AI legal compliance process & Evaluate and document how the AI system complies with relevant regulations and standards, identifying use case-specific legal risks and required controls. Apply the organization's legal compliance framework to ensure appropriate safeguards are in place, with clear documentation of compliance assessments and risk mitigations. \\ \hline
        CONTROL-035 & Establish societal impact assessment framework & Implement comprehensive processes for assessing and documenting potential societal impacts of AI systems, including effects on employment, economic systems, power dynamics, and cultural value. Include stakeholder consultation and impact mitigation planning. \\ \hline
        CONTROL-036 & Establish responsible development and deployment policy & Establish policies and procedures governing AI system development and deployment decisions that consider societal implications, including competitive pressures, governance gaps, and benefit distribution.  \\ \hline
        CONTROL-037 & Implement AI alignment validation system & Establish processes for validating and maintaining AI system alignment with human values and goals, including testing for goal preservation, monitoring for objective drift, and validation of decision-making processes against ethical standards. Includes specific attention to detecting and preventing potentially misaligned behaviors, emergent goals, or deceptive actions. Covers using interpretability techniques to measure and assure alignment with intended goals. \\ \hline
        CONTROL-038 & Establish AI Risk Management System & Implement a comprehensive AI risk management system including risk assessment processes, monitoring frameworks, governance structures, and response procedures.  \\ \hline
        CONTROL-039 & Establish data governance and management practices & Implement data governance measures used for training, including having a copyright policy and identifying and documenting data sources, potential biases, and mitigations taken.  \\ \hline
        CONTROL-040 & Establish documentation sharing mechanism & Implement a process to share information and documentation to third-parties, including to regulators and downstream deployers or developers. \\ \hline
        CONTROL-041 & Implement a risk reporting mechanism & Establish processes to identify and disclose known or reasonably foreseeable risks, the discovery of new risks, or instances of non-conformity to third parties.  \\ \hline
        CONTROL-042 & Establish a general purpose incident response mechanism & Establish processes to enable incident monitoring and reporting. This includes defining "serious incidents” or set a threshold for formal reporting based on regulatory requirements to third-parties, regulators, and impacted individuals. \\ 
\end{longtable}

\subsection{Colorado AI Act \citep{Colorado_AI_Act_2024} Policy Requirements}
\label{sec:appendix_requirements}
\begin{longtable}{| p{.17\textwidth} | p{.55\textwidth} | p{.13\textwidth}  | p{.15\textwidth} |} 
    \hline
        \rowcolor{darkgray}  
        \color{white} \textbf{Policy Req Key} & \color{white} \textbf{Policy Requirement} & \color{white} \textbf{Citation} & \color{white} \textbf{Related Control IDs}  \\    \hline
        CO-DEPLOYER.C1 & Deployers of high-risk AI systems must implement a risk management policy and program. & Section 6-1-1703.2 & CONTROL-012, CONTROL-014, \& CONTROL-038 \\ \hline
        CO-DEPLOYER.C2 & Deployers of high-risk AI systems must complete an impact assessment. & Section 6-1-1703.3 & CONTROL-035 \\ \hline
        CO-DEPLOYER.C3 & Deployers, or third parties contracted by the deployer, must review the deployment of the deployed high-risk AI system to ensure that the system is not causing algorithmic discrimination. & Section 6-1-1703.3 & CONTROL-015 \\ \hline
        CO-DEPLOYER.C4 & Deployers of high-risk AI systems must notify consumers subject to the high-risk AI system that the system will be deployed. & Section 6-1-1703.4 & CONTROL-025 \\ \hline
        CO-DEPLOYER.C5 & Deployers of high-risk AI systems that make consequential decisions adverse to consumers must provide an opportunity to appeal an adverse consequential decision. & Section 6-1-1703.4 & CONTROL-028 \\ \hline
        CO-DEPLOYER.C6 & Deployers must publish a statement summarizing the deployed high-risk AI systems. & Section 6-1-1703.5 & CONTROL-040 \\ \hline
        CO-DEPLOYER.C7 & Deployers of high-risk AI systems must disclose the discovery of algorithmic discrimination caused by a high-risk AI system. & Section 6-1-1703.7 & CONTROL-042 \\ \hline
        CO-DEPLOYER.C8 & AI systems intended to interact with consumers must disclose to consumers that they are interacting with an AI system. & Section 6-1-1704.1 & CONTROL-025 \\ \hline
        CO-DEVELOPER.C1 & Developers of high-risk AI systems must make available information and documentation to deployers or other developers of the high-risk AI system. & Section 6-1-1702.2 & CONTROL-040 \\ \hline
        CO-DEVELOPER.C2 & Developers must make available a statement summarizing the high-risk AI systems developed and how algorithmic discrimination risks are managed. & Section 6-1-1702.4 & CONTROL-040 \\ \hline
        CO-DEVELOPER.C3 & Developers of high-risk AI systems must disclose known or foreseeable risks of algorithmic discrimination caused by a high-risk AI system. & Section 6-1-1702.5 & CONTROL-041 \\ \hline
        CO-DEVELOPER.C4 & AI systems intended to interact with consumers must disclose to consumers that they are interacting with an AI system. & Section 6-1-1704.1 & CONTROL-025 \\ \hline
\end{longtable}

\end{document}